\documentclass[11pt,a4paper]{article}
\pdfoutput=1
\usepackage{jinstpub}
\usepackage[english]{babel}
\usepackage{upgreek}
\usepackage{xspace}
\usepackage{graphicx}
\usepackage[capitalize]{cleveref}
\usepackage{url}
\usepackage{paralist}
\usepackage{enumerate} 
\usepackage{enumitem}

\title{The FRAM robotic telescope for atmospheric monitoring at the Pierre Auger Observatory}
\author{\includegraphics[height=30mm]{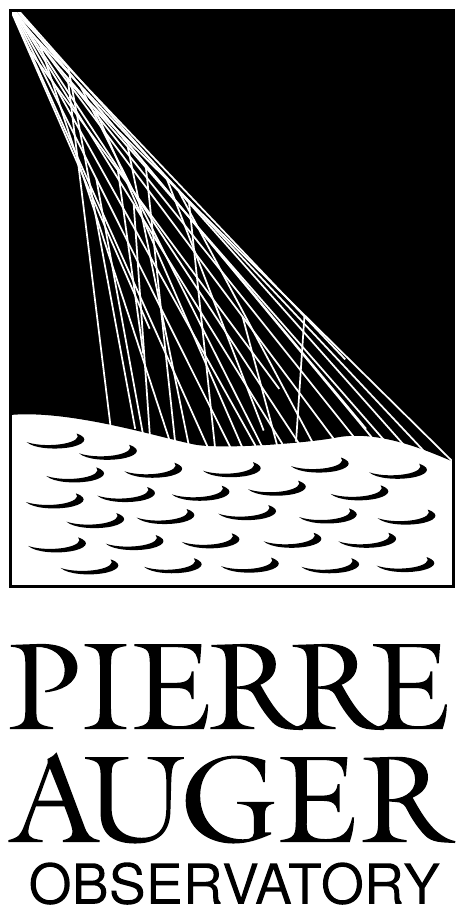}\\[3mm]The Pierre Auger Collaboration et al.}
\affiliation{Av.\ San Mart\'{\i}n Norte 306, 5613 Malarg\"ue, Mendoza, Argentina}
\emailAdd{auger\_spokespersons@fnal.gov}
\abstract{FRAM (F/Photometric Robotic Atmospheric Monitor) is a robotic telescope operated at the Pierre Auger Observatory in Argentina for the purposes of atmospheric monitoring using stellar photometry. As a passive system which does not produce any light that could interfere with the observations of the fluorescence telescopes of the observatory, it complements the active monitoring systems that use lasers. We discuss the applications of stellar photometry for atmospheric monitoring at optical observatories in general and the particular modes of operation employed by the Auger FRAM. We describe in detail the technical aspects of FRAM, the hardware and software requirements for a successful operation of a robotic telescope for such a purpose and their implementation within the FRAM system.}
\arxivnumber{2101.11602} 
\dedicated{\flushleft\textnormal{\textsc{%
Published in JINST as \href{http://dx.doi.org/10.1088/1748-0221/16/06/P06027}{DOI:10.1088/1748-0221/16/06/P06027}
\\
Report number: FERMILAB-PUB-21-025-AD-E-TD
}}}
\notoc
\begin{document}
\maketitle
\flushbottom

\section{Introduction}

The Pierre Auger Observatory \cite{auger}, located near Malargüe, Argentina, is currently the largest ultra-high energy cosmic ray observatory in the world. It measures the properties of extensive air showers induced by cosmic rays in the atmosphere. The observatory uses a hybrid detection technique and consists in its basic configuration of an array of surface particle detectors and five stations with optical telescopes overlooking this array from each side. These telescopes detect the faint fluorescence light emitted in the atmosphere due to the passage of the secondary particles of the extensive air showers. While the surface detectors work continuously, the fluorescence telescopes can be operated only in dark conditions during nights with low Moon illumination (approximately 15\% of the time). The measurement of the longitudinal profile of the fluorescence light by the fluorescence telescopes provides an estimate of the energy deposit of the showers applying the known air fluorescence yield and its dependence on local atmospheric conditions \cite{R1,R2} and the consideration of the so-called invisible energy \cite{escale}. The total energy of the primary cosmic ray is then determined as the sum of the calorimetric energy and the invisible energy. The longitudinal fluorescence profile also carries information about the character of the primary particle, and thus about the mass composition of the primary particle beam, and potentially also about the high-energy hadronic interactions that took place during the development of the shower.

In order to fully exploit the fluorescence technique, many aspects of atmospheric conditions must be continually monitored. The production yield of the fluorescence light depends on vertical profiles of temperature, humidity, and pressure. The transmission of this light is affected by scattering by molecules and aerosols (the contribution of absorption is negligible in the wavelength range used by the fluorescence telescopes) over distances up to tens of kilometres to the fluorescence telescopes. While the Rayleigh scattering by molecules can be determined from the state variables of the atmosphere, the Mie scattering by aerosols requires frequent dedicated measurements \cite{atmo1,astudy}. Using hourly vertical aerosol profiles, instead of an average profile, significantly improves the precision of the determination of both the energy of the primary particle and the depth of shower maximum \cite{harvey}. The presence of any clouds can also affect the determination of the aforementioned observables as well as of any other properties inferred from the shape of the profile (such as those used for particle physics studies), as the presence of the clouds can significantly distort the apparent profile. A series of cloud cuts from different instruments is applied to select events for physics analysis. For these reasons, the operation of fluorescence telescopes requires a sophisticated atmospheric monitoring system. 

\begin{figure}[!h]
\centering
\includegraphics[width=.6\textwidth]{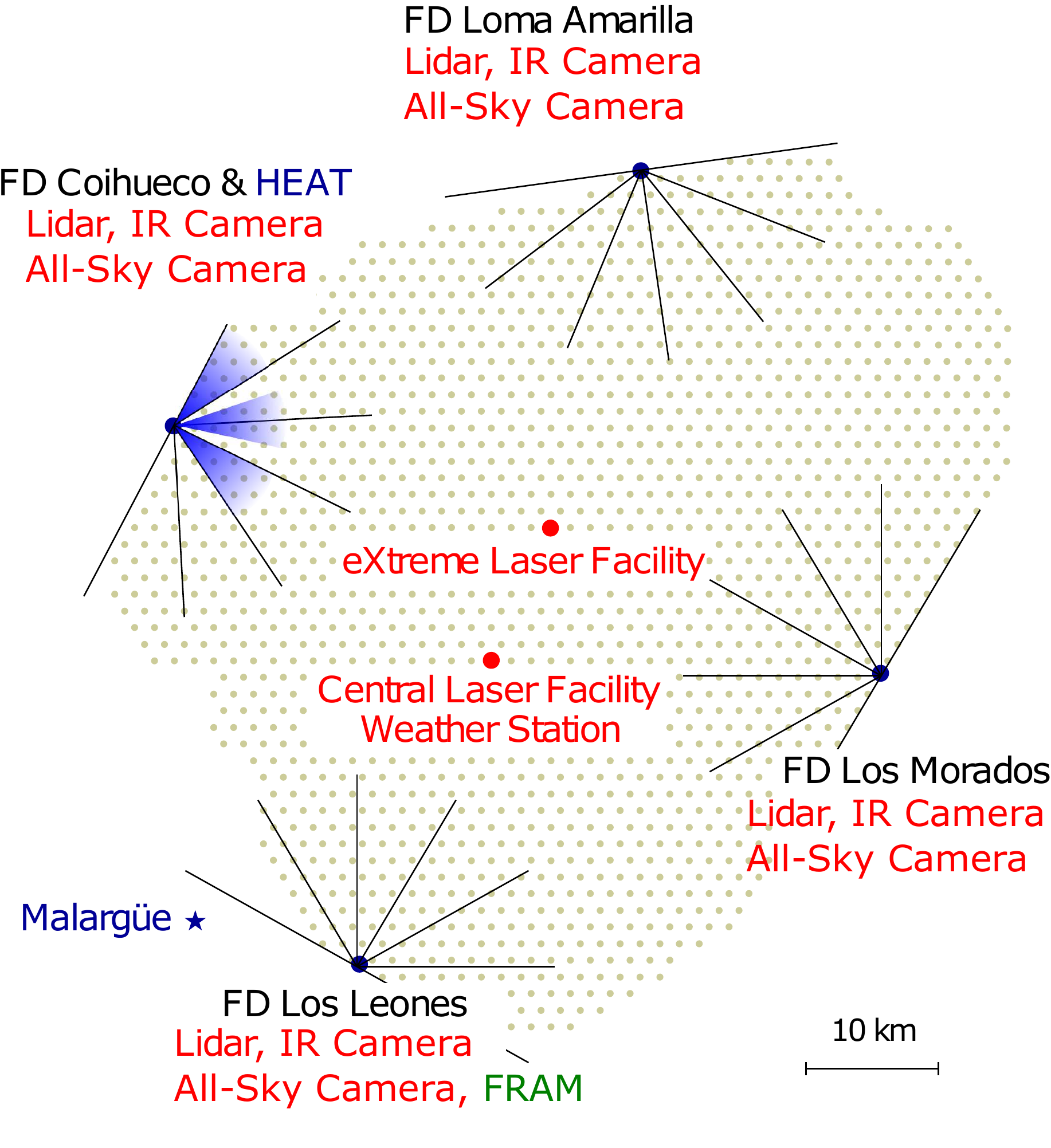}
\caption{\label{fig:map} Schematic overview of selected atmospheric monitoring devices installed at the Pierre Auger Observatory. At each FD site, there is, among other devices, a lidar station, an infrared camera for cloud cover detection and a visible-light All-Sky Camera. Two laser facilities (CLF and XLF) are installed close to the center of the surface detector array. The FRAM telescope is located at the Los Leones FD site (with a second FRAM planned for the Cohuieco FD site in the future).}
\end{figure}

At the Pierre Auger Observatory, the primary devices to assess the transparency of the atmosphere are two laser installations near the center of the array, called the CLF (Central Laser Facility) and the XLF (eXtreme Laser Facility) \cite{CLF}, used as reference light sources. The scattered light is then observed with the fluorescence telescopes themselves to infer the vertical profiles of aerosol extinction \cite{auger}. Additionally, a Raman lidar is operated at site of the CLF, sharing the laser beam with the CLF \cite{medina}. Furthermore, elastic lidars at the sites of the fluorescence sites primarily provide information on cloud cover and height \cite{lidar}; the cloud cover is also monitored using infra-red cloud cameras \cite{cloud} and visible-light All-Sky Cameras that detect clouds by star counting \cite{ASC}. The laser systems actively produce light which means that when used in the field of view of the fluorescence telescopes, the data taking is affected. The FRAM (F/Photometric Robotic Atmospheric Monitor) robotic telescope has been developed as an additional atmospheric monitoring device using stars as reference light sources. The first FRAM hardware was installed at the site of the first fluorescence telescope station (Los Leones) in 2005. Since then, the hardware, measurement methods, and applications of the system have gone through several major improvements. The layout of the observatory with the described devices is illustrated in Fig.~\ref{fig:map}.

Using stars as reference sources allows for determining the integral optical depth using a passive system. Indeed, the FRAM complements the laser-based methods where data are needed with high temporal and spacial resolution without interfering with the operation of the fluorescence telescopes. Currently the main application of FRAM at the Auger Observatory is the rapid monitoring of atmospheric conditions along the apparent path of showers that may have an anomalous longitudinal profile \cite{rapid}. Knowing actual atmospheric conditions for those showers enables us to exclude the possibility that such anomaly was caused by the presence of any clouds along the shower track ("Shoot-the-Shower") and hadronic interactions models can be tested and further constrained \cite{dblbmp}. The successful development of the Auger FRAM led to the proposal to include a similar instrument in the design of the Cherenkov Telescope Array (CTA), the future largest ground-based gamma-ray observatory in the world. The CTA FRAMs have a modified design tailored to the characteristics of the CTA observatory where they will monitor the changes in atmospheric transparency across the whole field of view of the Cherenkov telescopes simultaneously \cite{CTA}. Based on data from the wide-field telescopes installed at the future sites of the Cherenkov Telescope Array, it has been be shown that wide-field photometry can be used to measure the integral atmospheric extinction with a precision better than 0.01 optical depths \cite{cta_fram}. We plan to implement this method at the FRAM setup at the Pierre Auger Observatory in the near future.

\section{Applications of stellar photometry for atmospheric monitoring}

The basic idea of using stars as reference sources for atmospheric monitoring is simple: the difference between the observed brightness of a star and the predicted value based on the star’s known properties depends on the transparency of the atmosphere between the detector and the upper edge of the atmosphere in the given direction. More specifically, for each star observed, a model brightness is calculated from the catalogue value (in magnitudes) $m_\mathrm{cat}$ as 
\begin{equation}
m_\mathrm{model}=M\left(m_\mathrm{cat}+Z+f(C,x,y)+g(A,k,C)\right) \label{model},
\end{equation}
where $M$ accounts for possible non-linearity of the whole system (including the photometric method), $Z$ is the calibration constant of the system (so-called zeropoint), $f$ describes corrections due to the different response of the system to stars of different colors through some color index $C$ and in different parts of the field of view, and $g$ is a model of extinction depending on the airmass $A$, the color of the star and the extinction constant $k$, which is equal to the extinction measured at the zenith \cite{atmohead16}. The airmass $A$ is a dimensionless quantity expressing the integral mass of atmosphere (or a component thereof) encountered by the light of a star at a given altitude above the horizon relative to the same quantity at the zenith. If stars at different values of airmass are observed at once or in a short time window, the parameters of the instrument and of the atmosphere can be determined simultaneously as the former do not depend on the airmass while the latter do and can thus be separated using a fitting procedure, akin to the well-known Langley method \cite{langley}. As some instrumental parameters change only slowly in time, they may be fitted over a larger set of observations globally. In reality, both the observed and the predicted brightness are affected by uncertainties which determine the applicability of the method for different purposes and the optimal detection setup. 

The availability of precise photometric data for stars varies from star to star as there are many catalogs with varying levels of sky coverage. Moreover, typically a brightness value measured in a bandpass that does not match the spectral response of a given setup and thus data from several bandpasses are required for a reliable prediction of the brightness of the star. In standard photometric fields, there are many stars in a few small fields of the sky for which precise data in many wavelength bands exist, but using only these would severely limit the possibility to do measurements in an arbitrary direction. Furthermore, these isolated fields are typically unsuitable for any method that uses the Langley calibration, because the only way to observe them at various values of airmass is to wait for them to move due to the rotation of the Earth – however during this time, the conditions may change. For the purpose described here, it is thus better to use a suitable all-sky catalog, such as Tycho2 \cite{Tycho2}. Despite the precision of Tycho2 for individual stars being lower than that of dedicated photometric surveys, the homogeneity of its data over the entire sky is extremely valuable as it prevents the introduction of any biases when processing measurements of stars from large areas of the sky simultaneously. Using such a homogeneous catalog allows measurements of transparency in arbitrary directions, although the low precision for individual stars requires the observation of a large number of stars at once. The uncertainty of the catalog brightness in Tycho2 increases quickly above magnitudes of roughly 10. Since there are relatively few sufficiently bright stars in the sky, a wide-field setup (on the order of degrees) has to be preferred, although this property comes along with a relatively small aperture.

The uncertainty in the measured brightness of the stars depends on the specific instrumental setup. For a long time, photoelectric photometers were considered the gold standard in astronomical photometry thanks to their stability and absolute calibration, but they have been mostly superseded by CCD cameras for most applications. For a CCD camera, the uncertainty in a photometric measurement depends on the noise level and stability of the camera electronics and other hardware effects and then chiefly on the amount of light registered from the star. That in turn depends on exposure length and the aperture of the optics; for a given aperture, the maximal field of view is limited by practical limits of optics and thus there is always a trade-off between the precision of measurement for individual stars and the number of stars observed.

Based on these considerations and from practical experience, we have identified several basic operation modes in which a small robotic optical telescope can be employed in atmospheric monitoring for an optical observatory (not necessarily a set of fluorescence or Cherenkov telescopes).

\begin{figure}[!h]
\centering
\includegraphics[width=.9\textwidth]{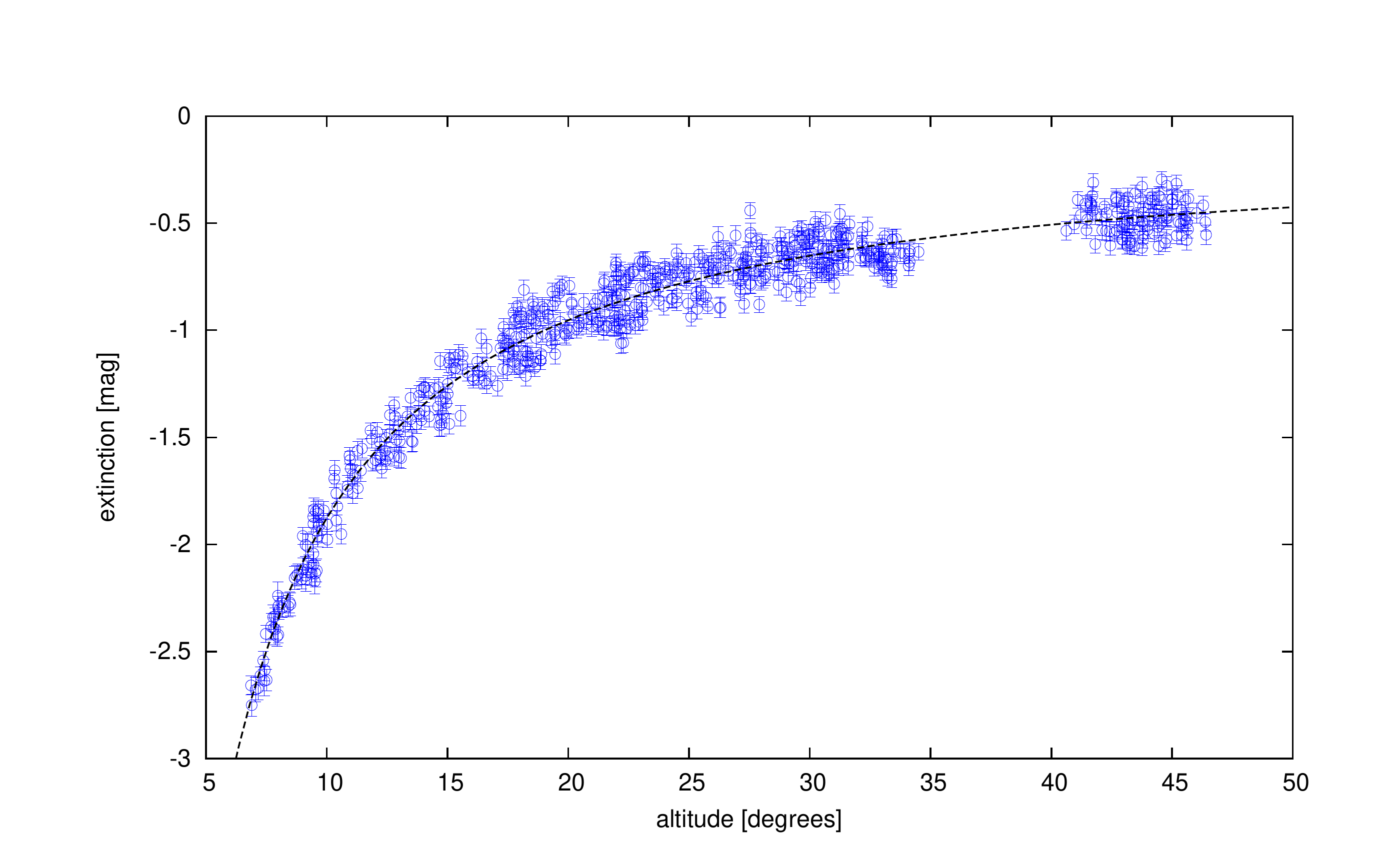}
\caption{\label{fig:extinct} An example of an altitude scan taken by the Auger FRAM. Each individual point represents a single star and the y-value is equal to the observed extinction for the star: the difference between its catalog and measured brightness, corrected for various instrumental and atmospheric effects. The fit of the extinction as a function of altitude (shown by the dashed line) allows the determination of the vertical aerosol optical depth. This particular scan has been originally taken in Mode B, triggered by a cosmic ray shower, but the same data can be used for precise aerosol measurements. In this mode, besides the continuous coverage of the field of view of the fluorescence detector between 1.5 and 31.5 degrees of altitude, also an image of the arrival direction of the shower is taken, which is also included in the fit to improve the lever arm. The error bars include the statistical and systematic uncertainties of both the measurement and the catalog values.}
\end{figure}

\paragraph{Mode A -- Precise aerosol measurement using altitude scans.}

\begin{figure}[!h]
\centering
\includegraphics[width=.9\textwidth]{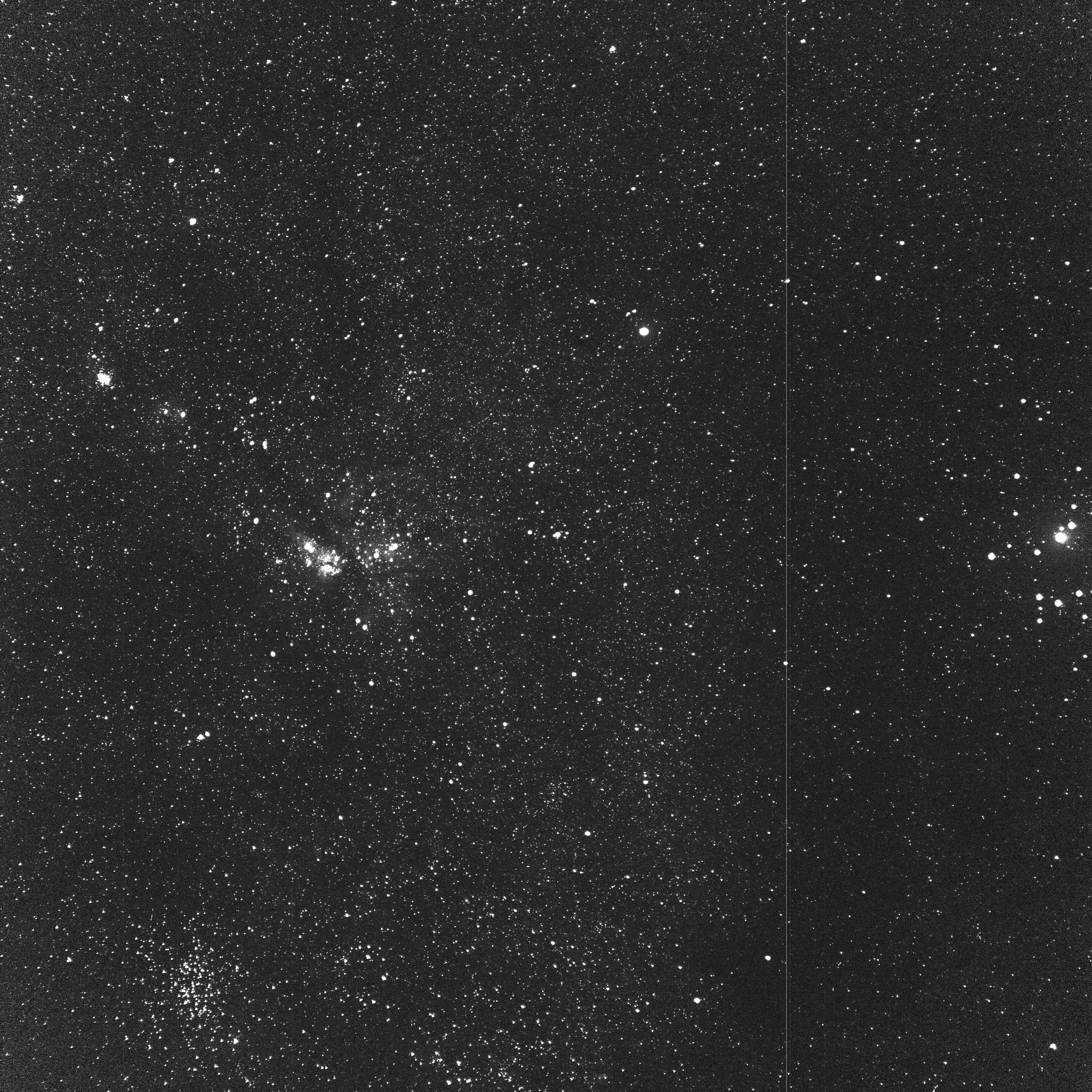}
\caption{\label{fig:field}  One of the images taken by the Auger FRAM during a Mode A altitude scan, in this case covering a well-known region of the Milky Way around the star $\eta$ Carinae and its surrounding nebula (seen left of the center). A single 30-second exposure in the Johnson B filter can be used to measure the brightness of hundreds of stars simultaneously in such a rich area.}
\end{figure}

Taking a series of images at different altitudes above the horizon and thus at different values of airmass $A$ allows for the simultaneous determination of the atmospheric extinction and the instrument calibration constant $Z$. This is achieved by a fit to the altitude dependence of the difference between predicted and observed brightness of the stars when assuming horizontal stratification of aerosols in the atmosphere as described in detail in \cite{cta_fram} -- see Fig.~\ref{fig:extinct}. This method provides the most precise value of the integral vertical aerosol optical depth, which is obtained from the shape of the altitude dependence and thus does not require prior knowledge of the calibration constant $Z$. In fact it can be used to provide calibration for measurements in other modes. However in order for the fit to work, the extinction must follow the assumed altitude dependence, which is violated in particular when clouds are present. In the presence of some clouds on the sky, the probability for successful measurement can be improved by using external data to select a cloud-free path for the scan on the sky; this is implemented on all the current FRAMs using data from the All-Sky Cameras that are present both at the Auger Observatory and at the future CTA sites. Potentially it can be also problematic when aerosols are distributed in a non-stratified manner. Such instances are clearly visible from the data as deviations from the fitted shape and can be excluded, so the result is not an invalid value, but no value at all. Therefore, in order to produce results, the method requires some level of homogeneity over several kilometers of distance from the telescope towards the azimuth of the scan, which is the typical distance where most of the observed extinction happens (usually only stars more than 7 degrees above the horizon are taken into account in altitude scans). However this is still a much smaller scale than that of the whole Auger Observatory ($65\times 45$ km$^2$). The method can also be used to investigate small inhomogeneities in the aerosol distribution by taking scans in different azimuth directions and this can also be repeated with high temporal resolution in order to quantify the changes in the aerosol conditions. At the Auger FRAM, the altitude scans are taken as a series of 30 s exposures taken with the photometric Johnson B filter (see Fig.~\ref{fig:field} for an example) at an azimuth based on the available real-time cloud data from other instruments and the position of the Moon in the sky. The Johnson B filter has been chosen for its proximity in wavelength to the near-UV region in which the Auger fluorescence telescopes operate, as photometry directly in this region is complicated by the spectral properties of the used optics (cf. Fig.~\ref{fig:spectral}) and the lack of suitable all-sky stellar catalogs for comparison. The measured extinction can be translated to the near-UV region using Eq.~\ref{angstrom}.

\paragraph{Mode B -- Triggered operation in a large field of view.}

\begin{figure}[!h]
\centering
\includegraphics[width=.9\textwidth]{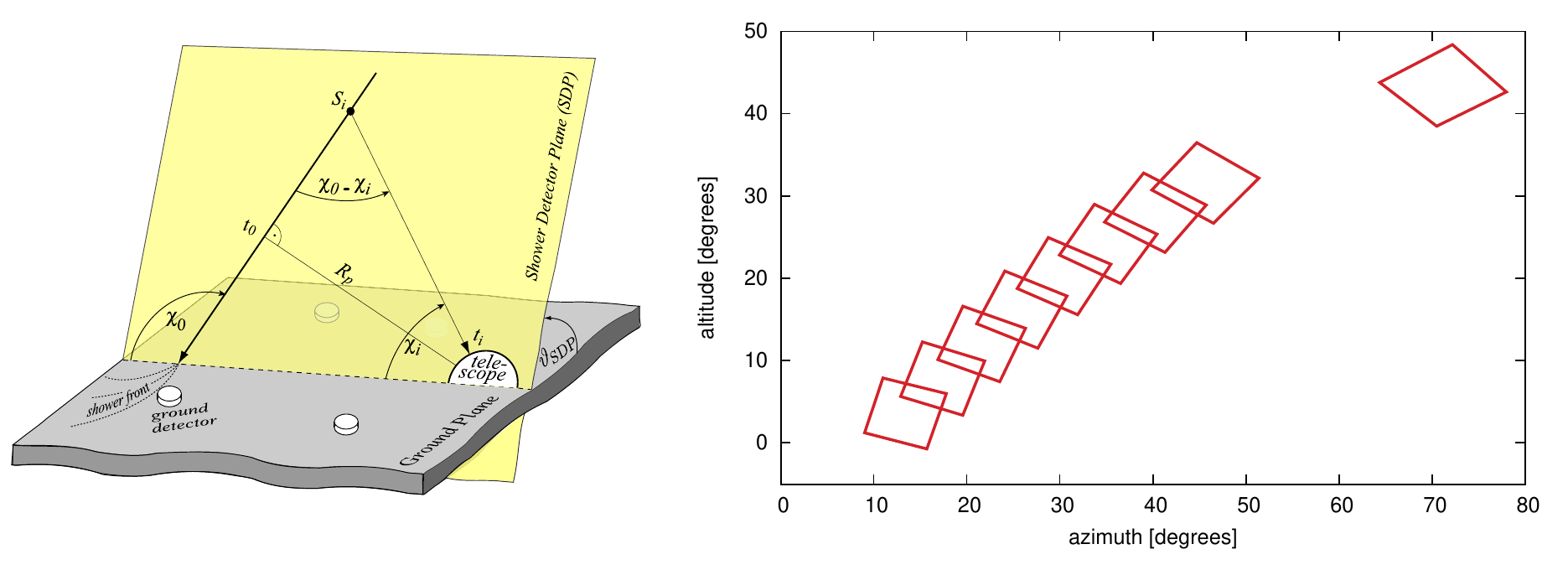}
\caption{\label{fig:fields} The apparent trajectory of an extensive air shower can be described by several geometric parameters (left) that are then passed to FRAM, where a series of observation fields (right) is generated in order to cover the trajectory across the whole FD field of view between 1.5 and 31.5 degrees of altitude as well as the direction of the arrival of the shower as a serendipity observation for the case that the cosmic ray was associated with a transient phenomenon. The observation depicted here is the same that produced the data shown in Fig.~\ref{fig:extinct}.}
\end{figure}

For a system such as the fluorescence detector of the Pierre Auger Observatory with a large field of view ($180^\circ \times 30^\circ$ for each of the four main FD telescope stations), it is not feasible to permanently monitor the whole field of view, but a robotic telescope can be used for timely determination of conditions in a selected part of the field of view. For triggering such a dedicated measurement, quasi-online reconstructed data from the fluorescence telescopes, indicating “interesting” events, can be used, based on a set of configurable cuts, as the rate of detection of showers by the fluorescence telescopes is too high to follow-up on all of them. When a shower is selected for observation by FRAM, the geometric parameters of its trajectory are passed to the FRAM system, which generates a set of fields to observe so that the trajectory is well covered, see Fig.~\ref{fig:fields}. The investigated trajectory then also forms a scan in altitude and thus the method of mode A can be applied directly to obtain a calibrated measurement. However if the aim is to identify inhomogeneities in the investigated region (such as in the case of the Shoot-the-Shower program at Auger), it is sufficient to just search for deviations from the theoretical altitude dependence of extinction and thus the method is extremely sensitive to any such disturbances as those are completely independent of the calibration of the system. On the other hand a horizontally uniform layer is completely undetectable through stellar photometry, but may be important for the subject at hand as an extensive air shower for example may pass through the layer making its light affected by the extinction in the layer only for a part of its trajectory. In principle, a similar method could be used for any other purpose where a suitable trigger can be found to define an area of interest within a larger field of view. If only deviations from the theoretical altitude dependence are of interest, the area can be of any shape; if precise aerosol measurements are needed, the area either needs to span a large range of airmass, or separate altitude scans (mode A) shall be made for calibration. 

\paragraph{Mode C -- Continuous monitoring of a small field of view.}

If the area of interest on the sky is sufficiently small so that it can be contained entirely within a field of view of a robotic telescope (such as is the case with the field of view of the CTA telescopes), it can be monitored continuously (with temporal resolution given by the length of exposure and readout) for changes in transparency. As in the previous case (Mode B), it is much easier to detect changes in transparency than to measure the absolute value of it in a single field. However even the latter is readily possible using the calibration provided by taking altitude scans (Mode A) at regular intervals, even though the precision of the extinction measured is smaller than that of the value obtained from a full scan, as it combines the uncertainty of the measurement in the field and that of the calibration determined from the scans. This is the operation mode foreseen for the CTA FRAMs during their future operation. The field of view itself will be cut into smaller areas of comparable numbers of stars using adaptive Voronoi tessellation in order to provide a 2D view of possible changes in the transparency across the field of view; when large changes are detected, the observation of the Cherenkov telescopes will be interrupted and the vertical profile will be assessed with a lidar \cite{ctacalib}. 

\paragraph{Mode D -- Measurement of wavelength dependence of aerosol extinction.}

The aerosol extinction is usually assumed to be inversely proportional to a power of the wavelength with an \AA ngstr\"om exponent $\alpha$ between 0 and 2, which varies due to changes in the physical composition of the aerosols, so that for the optical lengths $\tau_1$ and $\tau_2$ measured at wavelengths $\lambda_1$ and $\lambda_2$, we find
\begin{equation}
\frac{\tau_1}{\tau_2}=\left(\frac{\lambda_1}{\lambda_2}\right)^{-\alpha} .\label{angstrom}
\end{equation}

This can be determined in principle by taking altitude scans (Mode A) in several different wavebands (typically defined by photometric filters, such as the Johnson BVRI system used by both Auger and CTA FRAMs). Very precise values are needed to obtain a reasonable precision in $\alpha$ when the aerosol content of the atmosphere is small, as the absolute error of $\alpha$ is proportional to the relative errors of the individual measurements. For example when measuring at two mean wavelengths $\lambda_1$, $\lambda_2$, we receive
\begin{equation}
\Delta\alpha=\frac{\ln(1+\delta\tau_1)+\ln(1+\delta\tau_2)}{\ln\lambda_1-\ln\lambda_2}\approx\frac{\delta\tau_1+\delta\tau_2}{\ln\lambda_1-\ln\lambda_2} .
\end{equation}

Note that the effective wavelength of the measured extinction depends on the spectrum of the individual stars and the \AA ngstr\"om exponent itself; the former effect is accounted for within Eq.~(\ref{model}), where the dependence of $k$ on the color of the star is explicitly modeled, the latter can be dealt with in an iterative manner -- first processing the data while assuming e.g. $\alpha=1$ in Eq.~(\ref{model}), then using Eq.~(\ref{angstrom}) to improve the estimate of $\alpha$ and processing the data again.

For other modes of operation, the bandpass defined by the Johnson B filter (centered around 445\,nm \cite{binney}) can be used, in which the molecular contribution to the extinction is dominated by Rayleigh scattering which is easily calculated from the overall column density of the atmosphere as provided by e.g. a global model, such as the GDAS \cite{gdas}. For other Johnson filters (typically V, centered around 551\,nm and R, centered around 658\,nm), the contribution of molecular absorption mainly by water and ozone is both important and highly variable due to meteorological effects and thus must be carefully considered when subtracting the molecular component from the measured extinction in order to extract the aerosol optical depth. The Tycho2 catalog only has data in two filters, which are close to Johnson B and V which provide only a limited lever arm for the determination of $\alpha$ and thus other catalogs must be considered. For atmospheric monitoring at Auger and CTA, the effect of changing $\alpha$ is small as their primary calibration methods (lasers) operate at wavelengths close to those of the majority of the observed light – for example the contribution of the unknown wavelength dependence of the aerosol scattering to the uncertainty in shower energy is estimated to be only 0.5\% \cite{escale}. Nevertheless the knowledge of physical properties of aerosols may be of general interest \cite{ang1, ang2, ang3}. For the Auger FRAM, a measurement in this mode consists of a scan in altitude, during which a 30-second image is taken in each of the B, V and R filters for every field of the scan.

\begin{figure}[!h]
\centering
\includegraphics[width=.9\textwidth]{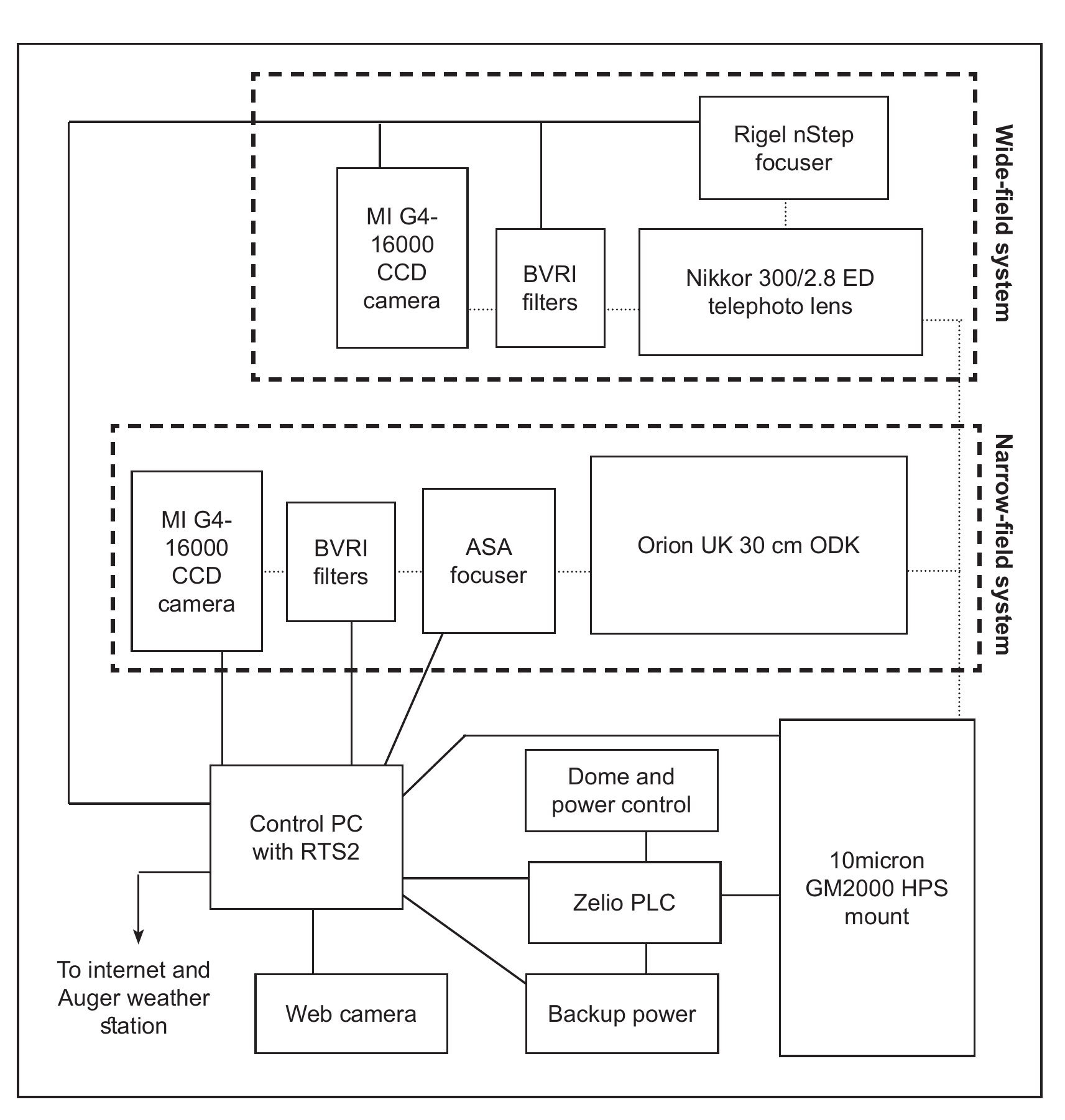}
\caption{\label{fig:schema} A simplified schema of the major components of the FRAM setup at the Pierre Auger Observatory. Lines between components indicate either mechanical binding (dotted) or electronic communication (solid).}
\end{figure}

\section{Hardware}

\subsection{General requirements}

Even though the Auger FRAM went through several major hardware changes, it has generally consisted of the same basic elements: a weatherproof enclosure, a German equatorial astronomical mount, two light detection systems (carried jointly by the same mount) – a small (20–30 cm) telescope first equipped with a photometer and later with a CCD camera (a “narrow-field”, NF, system) and a photographic lens with a CCD camera (a “wide-field", WF, system) – and a set of control and auxiliary devices. For a schematic overview of the system, see Fig.~\ref{fig:schema}.

The WF system allows both the measurement of a large number of stars for the purposes of precision aerosol measurement and the quick coverage of a large uninterrupted band of the sky for the purposes of the detection of clouds in the Shoot-the-Shower program and it is thus the primary tool used for atmospheric monitoring. The NF system currently provides the opportunity for additional astronomical observations (such as astrometry and/or photometry of asteroids, comets and variable stars, and gamma-ray burst follow-up \cite{malaga2}) within the spare time allowed by the atmospheric monitoring requirements, but it is also being developed for use in the atmospheric monitoring program. Such a possibility is particularly promising after the latest upgrade where the apparent field of view of the NF system has been considerably expanded, increasing the number of stars that can be measured at once.

One of the benefits of using stellar photometry for atmospheric monitoring is the possibility to assemble a highly capable device from affordable off-the-shelf products. The unique operating conditions of FRAM impose specific requirements on those devices for several reasons: 
\begin{inparaenum}[\itshape (i)]
\item The robotic nature of the telescope means that the observations are carried out continuously during every night, without breaks unless the weather conditions do not allow observations. Moreover, the exposures are typically short and in many observation modes, the target area on the sky is changed between each pair of exposures, resulting in many movements of the camera shutters and the mount.
\item The remote locations where astroparticle experiments (such as the Auger Observatory) are located require that the instruments are as autonomous as possible and most issues are solved remotely without local intervention. Regular on-site maintenance also should be kept to a minimum (typically once per year in the case of the Auger FRAM).
\item The environmental conditions at the sites are demanding, with considerable annual and diurnal changes in temperature, torrential rains, snow, periods of high humidity, but also of blowing dust. \end{inparaenum}
Additionally to the technical considerations, further requirements stem from the needs of the analysis of the atmospheric monitoring data and evolve with the progress of this analysis. 

The long experience with operating FRAM allows us to identify key issues, take appropriate steps to mitigate them, and select suitable products during hardware upgrades. We now describe these considerations for each part of the FRAM setup.

\begin{figure}[!h]
\centering
\includegraphics[width=.9\textwidth]{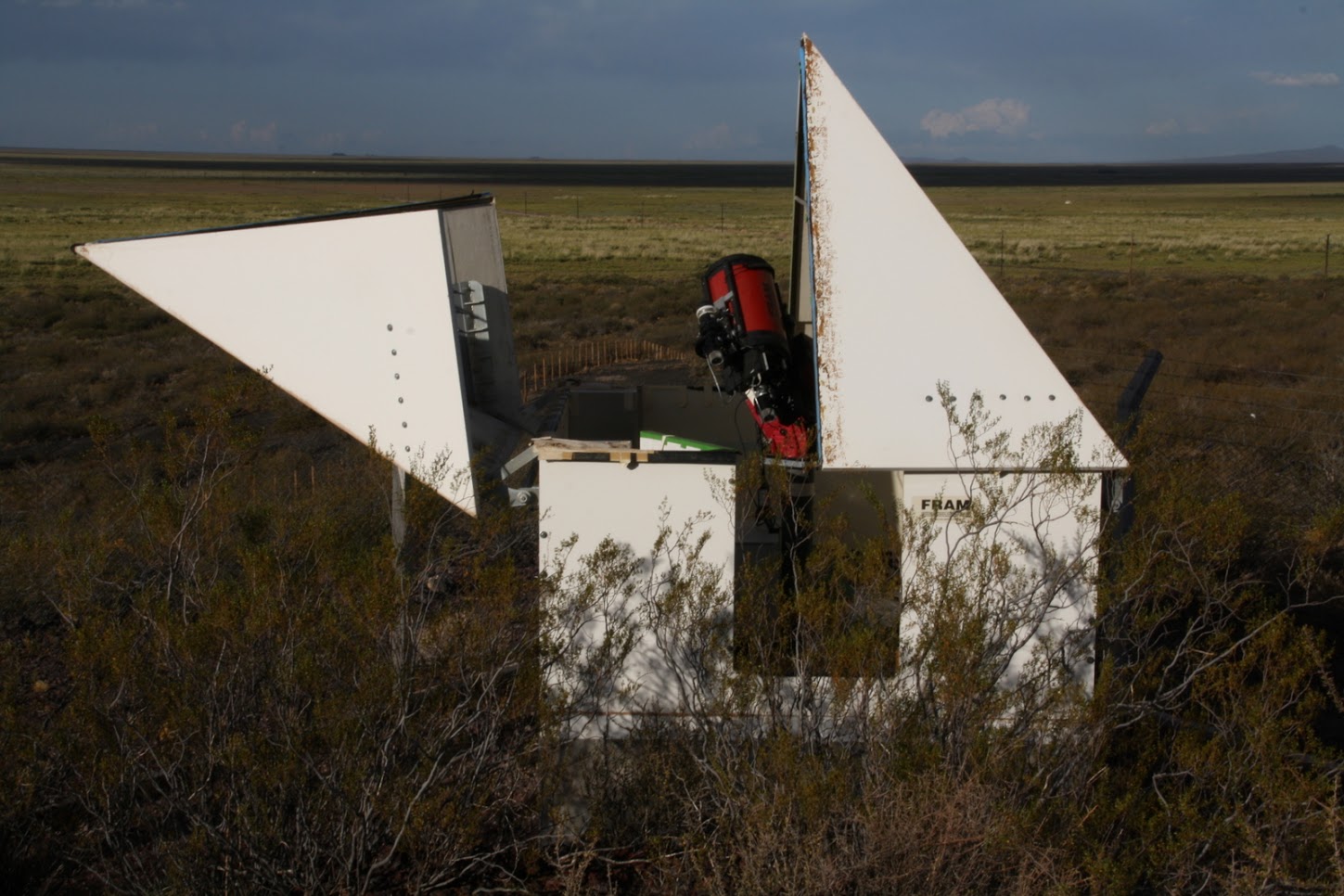}
\caption{\label{fig:dome} The FRAM enclosure with one half open and one closed. Inside is an older version of the FRAM setup with the Meade SCT and Paramount ME mount.}
\end{figure}

\subsection{Enclosure}

The enclosure protects the FRAM setup from the environment when it is not operating. As FRAM is expected to quickly react to triggers and to observe various parts of the sky in short sequence, a solution with a roof that fully opens to give access to the full sky at once is preferable to a more traditional rotating dome with a slit opening. In the FRAM case, the pyramidal roof consists of two independently operated halves (Fig.~\ref{fig:dome}), each moved by a pair of hydraulic cylinders. Mounted in the corners of the dome, the cylinders automatically balance the load (they are connected in parallel), and can open each half to almost 180°. The hydraulic system is designed for operation with considerable wind-forces acting on the dome halves in transition. The mechanical linkages are also designed with sufficient safety margin as they must hold the dome closed against the most extreme wind forces that might be experienced at the site of the Auger Observatory.
The hydraulic pressure is provided by a single pump driven by a single-phase AC motor. The whole system is now controlled by a Schneider Zelio smart-relay (PLC) module \cite{Schneider} using custom firmware. Since a custom-made board has proven to be not reliable enough, the industry-standard PLC solution has been implemented. 

The PLC has outputs that control the position of the hydraulic valves, the power to the pump motor, the source of this power (mains or backup) and the on/off state of the backup power inverter. Additionally, it controls the power to the mount, cameras and other devices to provide a reliable way of restarting them and it also provides a special output for the on/off control of the 10micron mount (see Sec.~\ref{mount}) For each half of the roof, mechanical end-switches are connected to the PLC inputs to indicate the position of a fully opened and fully closed roof. Further inputs provide the voltage of the (normally constantly charged) battery used to provide the 12~V DC voltage for the valves and the status of the mains AC power. 

The PLC communicates with the control PC via Modbus over TCP/IP. To achieve a fail-safe design to open the roof, the PC transmits a heartbeat signal (periodically changes a register value in the PLC). To close the roof, the PC simply stops the heartbeat. In case the communication with the PC is lost, the roof closes automatically, as this is simply equivalent to a command to close. 

On remote sites such as the Auger Observatory, a mains power cut is not uncommon. As the Auger Observatory does not provide global backup power to all its devices, and the Auger FRAM is fully dependent on this external power (unlike for example the solar-powered FRAMs at the southern CTA site), it has to be able to safely shut down in case of a power cut. Significant effort has been devoted to ensure the ability of the roof to close in such a situation. The pump motor creates a significant power spike on startup, which requires the use of a powerful inverter providing backup power from four lead-acid batteries. The inverter cannot be left powered during normal operation, because it causes electromagnetic interference in the CCD cameras, but it also must be given at least 30 seconds to stabilize after power on before the pump draws power from it – this is ensured through the logic in the PLC. During a power outage, the PLC runs from a UPS which also backs up the control PC, but the internal network is disconnected on purpose (the network switch is not backed up by the UPS) so that commands from the PC cannot interfere with the emergency closing of the roof. 

The hydraulic roof system is generally reliable, but regular maintenance (on the scale of once per year) is needed to check the status of the hydraulic connections and the oil level. The pump and all the hydraulic hoses had to be replaced after roughly 10 years of operation and a similar frequency of maintenance is needed in the future. The emergency closing system is a critical safety measure with several single points of failure and thus its integrity should be checked after any changes in the FRAM setup. 

\subsection{Mount}
\label{mount}

The mount points the cameras to the target area in the sky and follows the movement of the target due to the rotation of the Earth. For a remote robotic observatory like FRAM, the mount must be able to resume observation after a power cut without human intervention, which requires at least that it is able to find a “home” position autonomously – the lack of this ability has proven the original Losmandy G11 mount \cite{Losmandy} of FRAM as unsuitable. 

The replacement – Paramount ME \cite{Bisque} – had a simple optical sensor in each axis allowing it to return to its home position. All movements are evaluated relative to the home position from the movement of the stepper motors in each axis. After a power cut or mount restart, it must be ensured that the homing procedure is executed first before any attempt to reach a position on the sky, because commanding a move from a “desynchronized” state may cause movements of the mount beyond its physical boundaries. This does not harm the hardware, but requires human intervention, as reaching the physical stops prevents further autonomous movement of the mount in any direction. This is an intrinsic feature of the stepper motors used in Paramount ME, which need freedom of at least one step in both directions to properly initialize. A software daemon is installed at the control PC for ensuring the homing procedure. The Paramount ME required regular maintenance even beyond that prescribed by the vendor, due to the dusty environment of the Auger Observatory and the constant load of the robotic operation. Cleaning and greasing of the worm gear mechanisms had to be carried out at least once per year and the worms on both axes had to be replaced after five years of use. When the worms were in bad condition, the mount had a tendency to slip during movement and slew slightly off target – while this could be corrected using the astrometry of the star images, it occasionally led to the aforementioned problem of desynchronization and reaching the physical stops, requiring on-site intervention.

The current equatorial mount -- 10micron GM2000 HPS \cite{tenmicron} -- installed in September 2018, has absolute position sensors and its position is thus precisely known at any moment. Moreover, the manufacturer claims at least 10 years of maintenance-free operation. The GM2000 contains its own control computer, which needs to be booted and shut down properly. Normally this is done using a switch -- for remote operation, the switch is driven by one of the outputs of the roof PLC. The switch is stateless and there is no output from the mount to easily allow the PLC to determine whether it is on or off. Thus the control PC checks the state of the mount (over TCP/IP) continually and stores this information in a register of the PLC so that the PLC can autonomously handle the safe shutdown of the mount in case of a power cut. Since commissioning, the GM2000 has operated reliably and without maintenance. The internal communication between parts of the mount seems to be quite sensitive to electromagnetic disturbance and requires careful grounding. 

\subsection{Light sensors}

We have tested a photoelectric photometer as the main light detector, but we have not found the technology practical for the purposes of atmospheric monitoring: the measurement of a large number of stars is slow and even for individual stars, the reproducibility of the measurement was poor in the context of a remote robotic telescope due to issues with centering the star image on the aperture of the photometer automatically and various other issues that do not occur when an observer is physically present on site. On the other hand, CCD photometry easily allows simultaneous measurements of a large number of stars, even though it brings its own issues, especially regarding the stability of absolute calibration due to a multitude of effects including temperature dependences of the behavior of the camera electronics and focusing of the optics. With the combination of a careful data analysis and a well designed operational mode, the CCD cameras still give much better results than when using the photometer. CCD cameras are thus currently used in both Auger and CTA FRAMs. In the near future, we plan to test the performance in this application of CMOS detectors which, although superficially similar (as a large matrix of micron-sized pixels), are technologically very different from CCD cameras. The possibility to use SiPMs as detectors has been also considered but never explored further.

The Finger Lakes Instrumentation (FLI) \cite{FLI} cameras used in the first iterations of FRAM were not suitable for such a heavy use – their three-bladed iris shutters suffered from the repeated short exposures and became quickly unusable despite maintenance. Since 2012, all cameras used at FRAM are supplied by Moravian Instruments (MI) \cite{MI} and use a rotating shutter made of a single butterfly-shaped piece of metal which does not show any degradation in function even after years of use. Since one of the cameras failed without an apparent reason, the vendor now provides a special version of the cameras with protective coating on all electronics and the problem has not appeared again. However, care must be taken to regularly (at least yearly) replace the desiccant in the cold chamber as one of the CTA FRAM cameras was considerably damaged most likely due to corrosion caused by internal humidity. 

The most problematic part of the cameras is the filter wheel, which typically holds at least BVR Johnson photometric filters. For most applications, it would be sufficient to include a stationary B filter, but for the measurements of the \AA ngstr\"om exponent and for various astronomical applications, the ability to change filters is highly desirable. The MI filter wheel contains a large circular filter holder with a rubber band at the edge which is rotated by a small driving wheel; the positions of the filters are marked by holes in the holder that are detected by an optical gate, with the first position indicated by two holes as a “home” location. This setup has the tendency to become worn out and slip, causing a de-synchronization which then causes the applied filters to be shifted by one position from the requested ones. The vendor has delivered several improvements to the rubber band as well as detailed procedures for the setting of the pressure between the driving wheel and the rubber band, but the problem still occasionally appears. We have developed a software routine that checks for possible problems and restarts the filter wheel if needed. Most of the filter changes occur during the \AA ngstr\"om measurements (Mode D), which is always done as a series of scans with different filters – the procedure is such that in case of a filter wheel malfunction, it usually still obtains a series of consistent scans, just in a different set of filters and a large fraction of affected data can be salvaged automatically because from a full scan, the filter used can be relatively easily inferred.

\begin{figure}[!h]
\centering
\includegraphics[width=.9\textwidth]{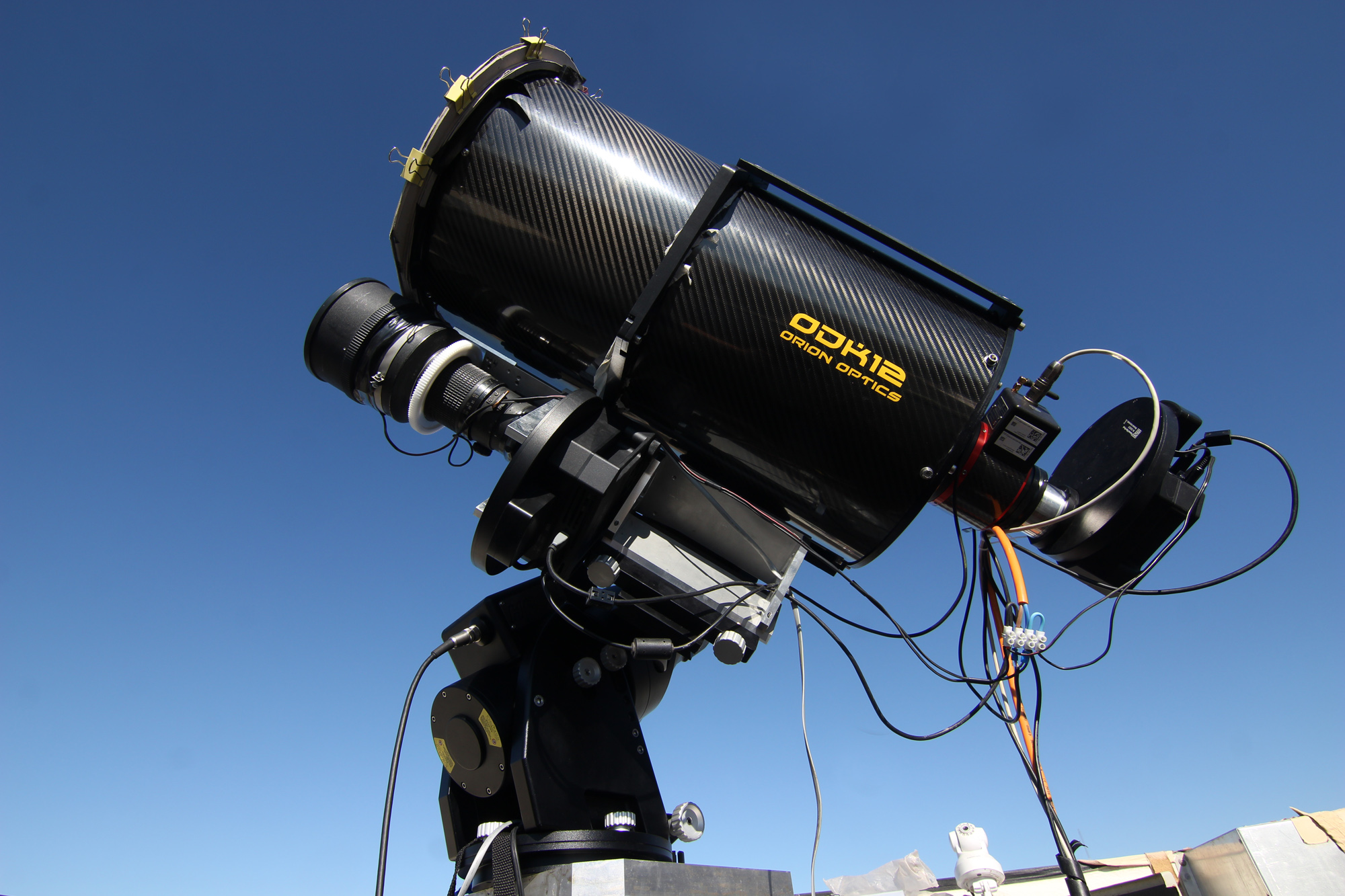}
\caption{\label{fig:telescopes} The current FRAM setup with the narrow-field system on the top and the wide-field system on the bottom.}
\end{figure}

\subsection{Wide-field system}

Originally using a Pentacon 200/2.8 lens with a small FLI camera and then a MI G2-1600, the wide-field system was significantly upgraded in 2013 to the current configuration of a Nikkor 300/2.8 lens and a large-format 36$\times$36 mm$^2$ MI G4-16000 camera (Fig.~\ref{fig:telescopes}). Even though the focal length of the Nikkor is longer, the large format camera covers an area of $7^\circ \times 7^\circ$, significantly larger than before the upgrade. The optical quality of the Nikkor lens degrades towards the corners of the image, as expected because the CCD chip is even larger than a traditional film frame for which the lens has been originally designed. The correction for this effect in Eq.~(\ref{model}) is acceptable in the whole field for the purposes of cloud detection in the Shoot-the-Shower program, but for precision aerosol measurements, only the inner circular part with a diameter of roughly 6.3 degrees, where this correction is less dependent on the momentary state of focus, is used; this still amounts to more than 60\% of the area of the CCD chip.

The fast f/2.8 lens has a narrow depth of focus. Moreover, it loses its focusing due to temperature changes and slippage due to the movement of the mount. Also due to residual color aberrations, the optimal focus position is slightly different for different filters, even though the filters themselves are confocal. Keeping proper focus is important in particular for operation in Mode C (monitoring of a single field) because changes in focusing have a strong effect on the calibration constant for stellar photometry as it affects the distribution of light around the center of the stellar image. This mode is not commonly employed at the Auger Observatory, but even for cloud detection or self-calibrated scans, it is necessary to have some means of controlling the focus as otherwise the lens eventually de-focuses to a useless state. The lens has only a manual focusing ring; to be able to focus it remotely, we thus use a focusing kit by Rigel Systems \cite{rigel} which consists of a plastic cogwheel attached to the focus ring of the lens which is driven by a USB-controlled stepper motor. 

Ideally, the lens should be aligned so that its optical axis intersects the CCD chip perpendicularly in the center of the chip as then the correction on the position of the star on the chip would be strictly radial, reducing the number of free parameters. The alignment can be checked easily by measuring the star shapes across the imaged field. A custom-made holding system for the lens and camera was designed and then gradually improved. Eventually such a task proves near impossible to perform flawlessly as there are many combined effects causing optical misalignment and not all of them are due to the mutual alignment of the camera and the lens. However it turns out that if a rough alignment is achieved, the remaining variations in star shape across the field of view can be satisfactorily treated during data processing, even if they are not radial, without introducing a bias in the measured atmospheric properties. Thus, further improvement is not required. 

\begin{figure}[!h]
\centering
\includegraphics[width=.9\textwidth]{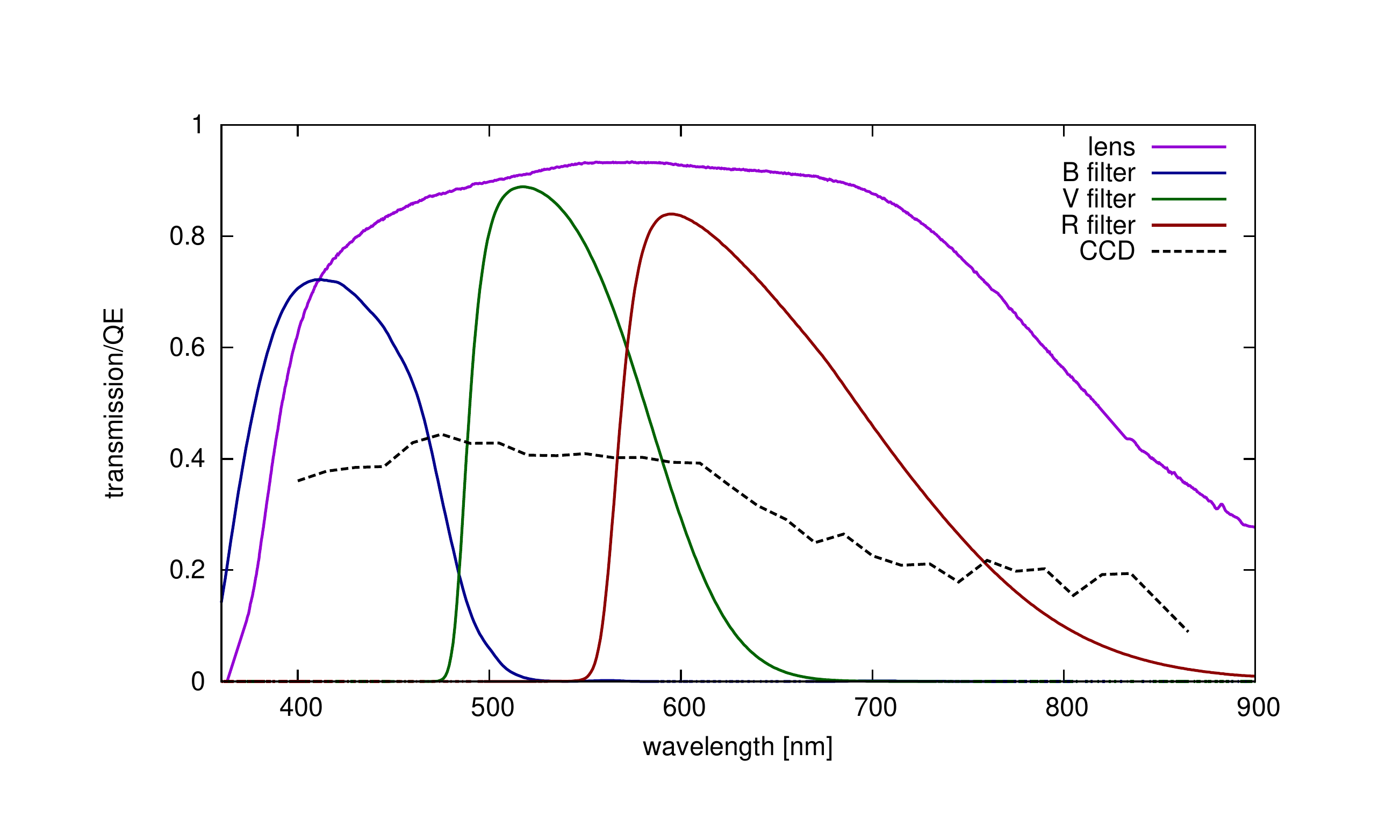}
\caption{\label{fig:spectral} Spectral characteristics of the key components of the wide-field system -- transmissivity of the lens and the filters and quantum efficiency of the CCD -- as measured in an optical laboratory.}
\end{figure}

Fig.~\ref{fig:spectral} shows the spectral characteristics of the key components of the wide-field system.

\subsection{Narrow-field system}

The narrow-field system is a standard small astronomical telescope. The original 20-centimeter Cassegrain telescope had significant problems with mechanical stability and has been replaced by a 30cm Meade \cite{Meade} Schmidt-Cassegrain (on loan from Instituto de Astrofisica Andalusia) and later by a 30-centimeter Orion UK \cite{Orion} Dall-Kirkham (Fig.~\ref{fig:telescopes}). The change to the Dall-Kirkham system was accompanied by a change from a small camera with a 14$\times$10 mm$^2$ chip to another G4-16000 with a 36$\times$36 mm$^2$ chip, which the Dall-Kirkham fully covers, albeit with significant vignetting; the setup achieves a full square degree of field of view. To avoid contributing to the vignetting, the system is equipped with a 3-inch focuser from Astro Systeme Austria \cite{ASA}.

\section{Control, software and operation}

All components of the FRAM system are controlled from a single PC running Linux. Apart from reliable basic components, a key consideration is the exclusive use of SSDs as spinning disks universally succumb to the environmental conditions on site. The requirements for local storage are substantial, as the Pierre Auger Observatory (as is typical for astroparticle experiments) is located in a remote area and transferring large amounts of low-priority data, such as the raw images from the CCD cameras, is not always easy. The PC is equipped with an IPMI interface allowing remote management and power control using a dedicated second ethernet interface. The most problematic aspect of the PC control lies in the USB communication with the various devices. Even though the problems could be solved by applying custom-made kernel patches, recently problems in USB communication appeared again, presumably because of the specific situation of a telescope setup, where many USB devices are mounted together in such a manner that their enclosures are conductively connected, possibly causing multiple ground loops. One possible solution, which is currently under investigation, could be the installation of optical isolators on the USB cables to prevent the ground loops and other sources of interference.

The whole FRAM operation is conducted within the RTS2 software framework, an open-source package for robotic observatories \cite{RTS2}. This modular framework was originally developed primarily for follow-ups of gamma-ray bursts (GRBs), but over time it has been extended for many purposes and is deployed on many observatories around the world. The Auger FRAM has been one of the test-beds for its development for years and some features of the package are tailored to the needs of the project \cite{RTS2b}. RTS2 can run an astronomical telescope with all its accessories (including the enclosure) autonomously. It is based on the elementary concept of a “target”, which can be an astronomical object described by its coordinates, or a more complicated recipe for observation defined via a Python interface using a specific API. During the night, the targets can be chosen from a database based on priorities assigned to them or the telescope set to follow nightly queues of planned observations in a given order. Thanks to the GRB background, RTS2 also naturally supports reactive observations triggered by external sources of information. For the Shoot-the-Shower program, a dedicated module has been developed to receive and process near real-time data from the Central Data Acquisition System of the Auger Observatory and take a decision to follow up an observation of a cosmic ray shower based on configurable sets of parameters.

Despite the autonomous capabilities of RTS2, the best results for FRAM in terms of uptime and data quality are still achieved with human supervision. Since 2012, we additionally rely on a dedicated observer, who checks the status of FRAM (and later also of the other FRAMs at CTA) every day before and sometimes during observations and resolves various issues or, if necessary, calls for on-site intervention. Over time, we have been able to identify the most common issues (such as race conditions among devices leading to data corruption, improper metadata recording, incorrect handling of the meridian flip of the mount etc.) and have either resolved them with improvements in hardware or by software workarounds, thus slowly eliminating the need for nightly human oversight. It is still not clear how to judge the quality of data taken automatically due to the variability of atmospheric conditions (how to determine whether any problems with data are due to unfavourable conditions or system issues) and thus regular checking of the outputs by an experienced observer is still desirable.

Apart from a multitude of bug fixes, the most common among the software solutions to the operational issues is the use of watchdogs over the drivers for the necessary devices. Each of the drivers runs as a separate process and in case of a crash, it is restarted, sometimes including an automatic power-cycle of the associated device. An important improvement is the development of a routine for automatic focusing, that can reliably decide if it is actually detecting stars and if the results are meaningful. It is implemented by analyzing a set of stars that are automatically detected in a sequence of images acquired with different focus positions. The images are cross-matched in order to reject spurious detections, and then a common minimum is found in the measured sequence of stellar FWHMs. If not enough star sequences are detected or if no clear minima are seen in their FWHMs, the routine properly identifies the focusing failure and does not update an optimal focus estimation, thus avoiding artificial focus drifts in bad or unstable weather conditions.

\section{Current FRAM setup and observation modes}

To summarize, the current FRAM setup at the Pierre Auger Observatory consists of a wide-field imager (MI G4-16000 CCD on a Nikkor 300/2.8 with a stepper motor external focuser) and a narrow-field imager (MI G4-16000 CCD on a 30-centimeter Orion UK Dall-Kirkham with 3” ASA focuser) jointly carried by a 10micron GM2000 HPS German equatorial mount, housed in a custom enclosure with a hydraulically opened roof controlled by a PLC and controlled from a PC running the RTS2 software with local modifications. 
The operation is mostly automated, but the status is checked almost nightly by a dedicated observer. The basis of the usual observation program is taking regular altitude scans (Mode A) taken as a series of 30 second exposures in the B filter in azimuths selected taking into account the data from the All-sky Camera and the position of the Moon (a too close passage to the Moon is avoided as such data are difficult to process properly); less frequently, scans are conducted in B, V and R filters for the \AA ngstr\"om coefficient measurements (Mode D). The scans are implemented using the Python API interface and a server-client configuration which allows multiple cameras (two in our case, NF and WF) to take data simultaneously and also allows clean resumption of scans in case of an interruption. In between these scans, some selected astronomical targets can be observed. The Shoot-the-Shower module is continually listening for triggers from the Central Data Acquisition System of the Auger Observatory and in case of an extensive air shower passing a set of predefined cuts, any ongoing observation is immediately stopped and the observation along the apparent path of the shower is commenced (Mode B). Regular runs of the auto-focusing routine are scheduled through the night to keep the focus consistent as the nightly decrease of outside temperatures causes contraction of the optical paths of the instruments.

\section{Conclusions}

Stellar photometry can be used for atmospheric monitoring in various modes, several of which are employed by the Auger FRAM. The Shoot-the-Shower program is the most valuable for further applications in physics analyses using the data of the Pierre Auger Observatory. Currently, the FRAM data on triggered air showers are being integrated into an analysis of cloud-free anomalous air shower events for studies of aspects of the mass composition of the primary beam and particle physics. The highest quality dataset comes from the latest WF setup installed in 2013, thus comprising more than 7 years so far. The expected number of truly anomalous events for the limited field of view of the fluorescence telescopes is still very low (at the order of 1\,\textperthousand \ of all detected events, according to the rough estimates of \cite{dblbmp}) and thus another FRAM installation is planned at the Cohuieco FD station. At that station, the low-energy extension HEAT \cite{HEAT} is also located, further increasing the number of detected showers. This second Auger FRAM will consist of two WF setups with identical optics, but different light sensors – alongside the traditional CCD camera, a CMOS detector will be tested to assess its viability for atmospheric monitoring. The methods to process the aerosol data from Mode A scans and \AA ngstr\"om coefficients from Mode D observations from both the existing FRAM and the future one are being finalized. All images taken by the Auger FRAM are calibrated and stored in a database. As part of the Open Data project at FZU - Institute of Physics of the Czech Academy of Sciences, all images were made publicly available through a web interface\footnote{\url{https://pc048b.fzu.cz/archive/}}, where they can be searched using various criteria for a range of astronomical applications, such as pre-discovery data on newly discovered Solar System objects or the study of specific variable stars. 


\section*{Acknowledgments}

\begin{sloppypar}
The successful installation, commissioning, and operation of the Pierre
Auger Observatory would not have been possible without the strong
commitment and effort from the technical and administrative staff in
Malarg\"ue. We are very grateful to the following agencies and
organizations for financial support:
\end{sloppypar}

\begin{sloppypar}
Argentina -- Comisi\'on Nacional de Energ\'\i{}a At\'omica; Agencia Nacional de
Promoci\'on Cient\'\i{}fica y Tecnol\'ogica (ANPCyT); Consejo Nacional de
Investigaciones Cient\'\i{}ficas y T\'ecnicas (CONICET); Gobierno de la
Provincia de Mendoza; Municipalidad de Malarg\"ue; NDM Holdings and Valle
Las Le\~nas; in gratitude for their continuing cooperation over land
access; Australia -- the Australian Research Council; Brazil -- Conselho
Nacional de Desenvolvimento Cient\'\i{}fico e Tecnol\'ogico (CNPq);
Financiadora de Estudos e Projetos (FINEP); Funda\c{c}\~ao de Amparo \`a
Pesquisa do Estado de Rio de Janeiro (FAPERJ); S\~ao Paulo Research
Foundation (FAPESP) Grants No.~2019/10151-2, No.~2010/07359-6 and
No.~1999/05404-3; Minist\'erio da Ci\^encia, Tecnologia, Inova\c{c}\~oes e
Comunica\c{c}\~oes (MCTIC); Czech Republic -- Grant No.~MSMT CR LTT18004,
LM2015038, LM2018102, CZ.02.1.01/0.0/0.0/16{\textunderscore}013/0001402,
CZ.02.1.01/0.0/0.0/18{\textunderscore}046/0016010 and
CZ.02.1.01/0.0/0.0/17{\textunderscore}049/0008422; France -- Centre de Calcul
IN2P3/CNRS; Centre National de la Recherche Scientifique (CNRS); Conseil
R\'egional Ile-de-France; D\'epartement Physique Nucl\'eaire et Corpusculaire
(PNC-IN2P3/CNRS); D\'epartement Sciences de l'Univers (SDU-INSU/CNRS);
Institut Lagrange de Paris (ILP) Grant No.~LABEX ANR-10-LABX-63 within
the Investissements d'Avenir Programme Grant No.~ANR-11-IDEX-0004-02;
Germany -- Bundesministerium f\"ur Bildung und Forschung (BMBF); Deutsche
Forschungsgemeinschaft (DFG); Finanzministerium Baden-W\"urttemberg;
Helmholtz Alliance for Astroparticle Physics (HAP);
Helmholtz-Gemeinschaft Deutscher Forschungszentren (HGF); Ministerium
f\"ur Innovation, Wissenschaft und Forschung des Landes
Nordrhein-Westfalen; Ministerium f\"ur Wissenschaft, Forschung und Kunst
des Landes Baden-W\"urttemberg; Italy -- Istituto Nazionale di Fisica
Nucleare (INFN); Istituto Nazionale di Astrofisica (INAF); Ministero
dell'Istruzione, dell'Universit\'a e della Ricerca (MIUR); CETEMPS Center
of Excellence; Ministero degli Affari Esteri (MAE); M\'exico -- Consejo
Nacional de Ciencia y Tecnolog\'\i{}a (CONACYT) No.~167733; Universidad
Nacional Aut\'onoma de M\'exico (UNAM); PAPIIT DGAPA-UNAM; The Netherlands
-- Ministry of Education, Culture and Science; Netherlands Organisation
for Scientific Research (NWO); Dutch national e-infrastructure with the
support of SURF Cooperative; Poland -Ministry of Science and Higher
Education, grant No.~DIR/WK/2018/11; National Science Centre, Grants
No.~2013/08/M/ST9/00322, No.~2016/23/B/ST9/01635 and No.~HARMONIA
5--2013/10/M/ST9/00062, UMO-2016/22/M/ST9/00198; Portugal -- Portuguese
national funds and FEDER funds within Programa Operacional Factores de
Competitividade through Funda\c{c}\~ao para a Ci\^encia e a Tecnologia
(COMPETE); Romania -- Romanian Ministry of Education and Research, the
Program Nucleu within MCI (PN19150201/16N/2019 and PN19060102) and
project PN-III-P1-1.2-PCCDI-2017-0839/19PCCDI/2018 within PNCDI III;
Slovenia -- Slovenian Research Agency, grants P1-0031, P1-0385, I0-0033,
N1-0111; Spain -- Ministerio de Econom\'\i{}a, Industria y Competitividad
(FPA2017-85114-P and PID2019-104676GB-C32, Xunta de Galicia (ED431C
2017/07), Junta de Andaluc\'\i{}a (SOMM17/6104/UGR, P18-FR-4314) Feder Funds,
RENATA Red Nacional Tem\'atica de Astropart\'\i{}culas (FPA2015-68783-REDT) and
Mar\'\i{}a de Maeztu Unit of Excellence (MDM-2016-0692); USA -- Department of
Energy, Contracts No.~DE-AC02-07CH11359, No.~DE-FR02-04ER41300,
No.~DE-FG02-99ER41107 and No.~DE-SC0011689; National Science Foundation,
Grant No.~0450696; The Grainger Foundation; Marie Curie-IRSES/EPLANET;
European Particle Physics Latin American Network; and UNESCO.
\end{sloppypar}

\bibliographystyle{JHEP}
\bibliography{auger-fram}

\begin{center}
\rule{0.1\columnwidth}{0.5pt}\,\raisebox{-0.5pt}{\rule{0.05\columnwidth}{1.5pt}}~\raisebox{-0.375ex}{\scriptsize$\bullet$}~\raisebox{-0.5pt}{\rule{0.05\columnwidth}{1.5pt}}\,\rule{0.1\columnwidth}{0.5pt}
\end{center}

\section*{The Pierre Auger Collaboration}

A.~Aab$^{80}$,
P.~Abreu$^{72}$,
M.~Aglietta$^{52,50}$,
J.M.~Albury$^{12}$,
I.~Allekotte$^{1}$,
A.~Almela$^{8,11}$,
J.~Alvarez-Mu\~niz$^{79}$,
R.~Alves Batista$^{80}$,
G.A.~Anastasi$^{61,50}$,
L.~Anchordoqui$^{87}$
% ...

{\footnotesize

\begin{description}[labelsep=0.2em,align=right,labelwidth=0.7em,labelindent=0em,leftmargin=2em,noitemsep]
\item[$^{1}$] Centro At\'omico Bariloche and Instituto Balseiro (CNEA-UNCuyo-CONICET), San Carlos de Bariloche, Argentina
\item[$^{2}$] Centro de Investigaciones en L\'aseres y Aplicaciones, CITEDEF and CONICET, Villa Martelli, Argentina
\item[$^{3}$] Departamento de F\'\i{}sica and Departamento de Ciencias de la Atm\'osfera y los Oc\'eanos, FCEyN, Universidad de Buenos Aires and CONICET, Buenos Aires, Argentina
\item[$^{4}$] IFLP, Universidad Nacional de La Plata and CONICET, La Plata, Argentina
\item[$^{5}$] Instituto de Astronom\'\i{}a y F\'\i{}sica del Espacio (IAFE, CONICET-UBA), Buenos Aires, Argentina
\item[$^{6}$] Instituto de F\'\i{}sica de Rosario (IFIR) -- CONICET/U.N.R.\ and Facultad de Ciencias Bioqu\'\i{}micas y Farmac\'euticas U.N.R., Rosario, Argentina
\item[$^{7}$] Instituto de Tecnolog\'\i{}as en Detecci\'on y Astropart\'\i{}culas (CNEA, CONICET, UNSAM), and Universidad Tecnol\'ogica Nacional -- Facultad Regional Mendoza (CONICET/CNEA), Mendoza, Argentina
\item[$^{8}$] Instituto de Tecnolog\'\i{}as en Detecci\'on y Astropart\'\i{}culas (CNEA, CONICET, UNSAM), Buenos Aires, Argentina
\item[$^{9}$] Observatorio Pierre Auger, Malarg\"ue, Argentina
\item[$^{10}$] Observatorio Pierre Auger and Comisi\'on Nacional de Energ\'\i{}a At\'omica, Malarg\"ue, Argentina
\item[$^{11}$] Universidad Tecnol\'ogica Nacional -- Facultad Regional Buenos Aires, Buenos Aires, Argentina
\item[$^{12}$] University of Adelaide, Adelaide, S.A., Australia
\item[$^{13}$] Universit\'e Libre de Bruxelles (ULB), Brussels, Belgium
\item[$^{14}$] Vrije Universiteit Brussels, Brussels, Belgium
\item[$^{15}$] Centro Brasileiro de Pesquisas Fisicas, Rio de Janeiro, RJ, Brazil
\item[$^{16}$] Centro Federal de Educa\c{c}\~ao Tecnol\'ogica Celso Suckow da Fonseca, Nova Friburgo, Brazil
\item[$^{17}$] Universidade de S\~ao Paulo, Escola de Engenharia de Lorena, Lorena, SP, Brazil
\item[$^{18}$] Universidade de S\~ao Paulo, Instituto de F\'\i{}sica de S\~ao Carlos, S\~ao Carlos, SP, Brazil
\item[$^{19}$] Universidade de S\~ao Paulo, Instituto de F\'\i{}sica, S\~ao Paulo, SP, Brazil
\item[$^{20}$] Universidade Estadual de Campinas, IFGW, Campinas, SP, Brazil
\item[$^{21}$] Universidade Estadual de Feira de Santana, Feira de Santana, Brazil
\item[$^{22}$] Universidade Federal do ABC, Santo Andr\'e, SP, Brazil
\item[$^{23}$] Universidade Federal do Paran\'a, Setor Palotina, Palotina, Brazil
\item[$^{24}$] Universidade Federal do Rio de Janeiro, Instituto de F\'\i{}sica, Rio de Janeiro, RJ, Brazil
\item[$^{25}$] Universidade Federal do Rio de Janeiro (UFRJ), Observat\'orio do Valongo, Rio de Janeiro, RJ, Brazil
\item[$^{26}$] Universidade Federal Fluminense, EEIMVR, Volta Redonda, RJ, Brazil
\item[$^{27}$] Universidad de Medell\'\i{}n, Medell\'\i{}n, Colombia
\item[$^{28}$] Universidad Industrial de Santander, Bucaramanga, Colombia
\item[$^{29}$] Charles University, Faculty of Mathematics and Physics, Institute of Particle and Nuclear Physics, Prague, Czech Republic
\item[$^{30}$] Institute of Physics of the Czech Academy of Sciences, Prague, Czech Republic
\item[$^{31}$] Palacky University, RCPTM, Olomouc, Czech Republic
\item[$^{32}$] CNRS/IN2P3, IJCLab, Universit\'e Paris-Saclay, Orsay, France
\item[$^{33}$] Laboratoire de Physique Nucl\'eaire et de Hautes Energies (LPNHE), Sorbonne Universit\'e, Universit\'e de Paris, CNRS-IN2P3, Paris, France
\item[$^{34}$] Univ.\ Grenoble Alpes, CNRS, Grenoble Institute of Engineering Univ.\ Grenoble Alpes, LPSC-IN2P3, 38000 Grenoble, France
\item[$^{35}$] Universit\'e Paris-Saclay, CNRS/IN2P3, IJCLab, Orsay, France
\item[$^{36}$] Bergische Universit\"at Wuppertal, Department of Physics, Wuppertal, Germany
\item[$^{37}$] Karlsruhe Institute of Technology (KIT), Institute for Experimental Particle Physics, Karlsruhe, Germany
\item[$^{38}$] Karlsruhe Institute of Technology (KIT), Institut f\"ur Prozessdatenverarbeitung und Elektronik, Karlsruhe, Germany
\item[$^{39}$] Karlsruhe Institute of Technology (KIT), Institute for Astroparticle Physics, Karlsruhe, Germany
\item[$^{40}$] RWTH Aachen University, III.\ Physikalisches Institut A, Aachen, Germany
\item[$^{41}$] Universit\"at Hamburg, II.\ Institut f\"ur Theoretische Physik, Hamburg, Germany
\item[$^{42}$] Universit\"at Siegen, Department Physik -- Experimentelle Teilchenphysik, Siegen, Germany
\item[$^{43}$] Gran Sasso Science Institute, L'Aquila, Italy
\item[$^{44}$] INFN Laboratori Nazionali del Gran Sasso, Assergi (L'Aquila), Italy
\item[$^{45}$] INFN, Sezione di Catania, Catania, Italy
\item[$^{46}$] INFN, Sezione di Lecce, Lecce, Italy
\item[$^{47}$] INFN, Sezione di Milano, Milano, Italy
\item[$^{48}$] INFN, Sezione di Napoli, Napoli, Italy
\item[$^{49}$] INFN, Sezione di Roma ``Tor Vergata'', Roma, Italy
\item[$^{50}$] INFN, Sezione di Torino, Torino, Italy
\item[$^{51}$] Istituto di Astrofisica Spaziale e Fisica Cosmica di Palermo (INAF), Palermo, Italy
\item[$^{52}$] Osservatorio Astrofisico di Torino (INAF), Torino, Italy
\item[$^{53}$] Politecnico di Milano, Dipartimento di Scienze e Tecnologie Aerospaziali , Milano, Italy
\item[$^{54}$] Universit\`a del Salento, Dipartimento di Matematica e Fisica ``E.\ De Giorgi'', Lecce, Italy
\item[$^{55}$] Universit\`a dell'Aquila, Dipartimento di Scienze Fisiche e Chimiche, L'Aquila, Italy
\item[$^{56}$] Universit\`a di Catania, Dipartimento di Fisica e Astronomia, Catania, Italy
\item[$^{57}$] Universit\`a di Milano, Dipartimento di Fisica, Milano, Italy
\item[$^{58}$] Universit\`a di Napoli ``Federico II'', Dipartimento di Fisica ``Ettore Pancini'', Napoli, Italy
\item[$^{59}$] Universit\`a di Palermo, Dipartimento di Fisica e Chimica ''E.\ Segr\`e'', Palermo, Italy
\item[$^{60}$] Universit\`a di Roma ``Tor Vergata'', Dipartimento di Fisica, Roma, Italy
\item[$^{61}$] Universit\`a Torino, Dipartimento di Fisica, Torino, Italy
\item[$^{62}$] Benem\'erita Universidad Aut\'onoma de Puebla, Puebla, M\'exico
\item[$^{63}$] Centro de Investigaci\'on y de Estudios Avanzados del IPN (CINVESTAV), M\'exico, D.F., M\'exico
\item[$^{64}$] Unidad Profesional Interdisciplinaria en Ingenier\'\i{}a y Tecnolog\'\i{}as Avanzadas del Instituto Polit\'ecnico Nacional (UPIITA-IPN), M\'exico, D.F., M\'exico
\item[$^{65}$] Universidad Aut\'onoma de Chiapas, Tuxtla Guti\'errez, Chiapas, M\'exico
\item[$^{66}$] Universidad Michoacana de San Nicol\'as de Hidalgo, Morelia, Michoac\'an, M\'exico
\item[$^{67}$] Universidad Nacional Aut\'onoma de M\'exico, M\'exico, D.F., M\'exico
\item[$^{68}$] Universidad Nacional de San Agustin de Arequipa, Facultad de Ciencias Naturales y Formales, Arequipa, Peru
\item[$^{69}$] Institute of Nuclear Physics PAN, Krakow, Poland
\item[$^{70}$] University of \L{}\'od\'z, Faculty of Astrophysics, \L{}\'od\'z, Poland
\item[$^{71}$] University of \L{}\'od\'z, Faculty of High-Energy Astrophysics,\L{}\'od\'z, Poland
\item[$^{72}$] Laborat\'orio de Instrumenta\c{c}\~ao e F\'\i{}sica Experimental de Part\'\i{}culas -- LIP and Instituto Superior T\'ecnico -- IST, Universidade de Lisboa -- UL, Lisboa, Portugal
\item[$^{73}$] ``Horia Hulubei'' National Institute for Physics and Nuclear Engineering, Bucharest-Magurele, Romania
\item[$^{74}$] Institute of Space Science, Bucharest-Magurele, Romania
\item[$^{75}$] University Politehnica of Bucharest, Bucharest, Romania
\item[$^{76}$] Center for Astrophysics and Cosmology (CAC), University of Nova Gorica, Nova Gorica, Slovenia
\item[$^{77}$] Experimental Particle Physics Department, J.\ Stefan Institute, Ljubljana, Slovenia
\item[$^{78}$] Universidad de Granada and C.A.F.P.E., Granada, Spain
\item[$^{79}$] Instituto Galego de F\'\i{}sica de Altas Enerx\'\i{}as (IGFAE), Universidade de Santiago de Compostela, Santiago de Compostela, Spain
\item[$^{80}$] IMAPP, Radboud University Nijmegen, Nijmegen, The Netherlands
\item[$^{81}$] KVI -- Center for Advanced Radiation Technology, University of Groningen, Groningen, The Netherlands
\item[$^{82}$] Nationaal Instituut voor Kernfysica en Hoge Energie Fysica (NIKHEF), Science Park, Amsterdam, The Netherlands
\item[$^{83}$] Stichting Astronomisch Onderzoek in Nederland (ASTRON), Dwingeloo, The Netherlands
\item[$^{84}$] Universiteit van Amsterdam, Faculty of Science, Amsterdam, The Netherlands
\item[$^{85}$] Case Western Reserve University, Cleveland, OH, USA
\item[$^{86}$] Colorado School of Mines, Golden, CO, USA
\item[$^{87}$] Department of Physics and Astronomy, Lehman College, City University of New York, Bronx, NY, USA
\item[$^{88}$] Louisiana State University, Baton Rouge, LA, USA
\item[$^{89}$] Michigan Technological University, Houghton, MI, USA
\item[$^{90}$] New York University, New York, NY, USA
\item[$^{91}$] Pennsylvania State University, University Park, PA, USA
\item[$^{92}$] University of Chicago, Enrico Fermi Institute, Chicago, IL, USA
\item[$^{93}$] University of Delaware, Department of Physics and Astronomy, Bartol Research Institute, Newark, DE, USA
\item[$^{94}$] University of Wisconsin-Madison, Department of Physics and WIPAC, Madison, WI, USA
\item[] -----
\item[$^{a}$] Fermi National Accelerator Laboratory, Fermilab, Batavia, IL, USA
\item[$^{b}$] Max-Planck-Institut f\"ur Radioastronomie, Bonn, Germany
\item[$^{c}$] School of Physics and Astronomy, University of Leeds, Leeds, United Kingdom
\item[$^{d}$] Colorado State University, Fort Collins, CO, USA
\item[$^{e}$] now at Hakubi Center for Advanced Research and Graduate School of Science, Kyoto University, Kyoto, Japan
\item[$^{f}$] also at University of Bucharest, Physics Department, Bucharest, Romania
\item[$^{g}$] Nicolaus Copernicus Astronomical Center, Polish Academy of Sciences, Warsaw, Poland
\item[$^{h}$] Astronomical Institute of the Czech Academy of Sciences, Ond\v{r}ejov, Czech Republic
\item[$^{i}$] Vera C. Rubin Observatory, La Serena, Chile 
\end{description}

}

\end{document}